\documentclass[letterpaper]{article}
\usepackage[utf8]{inputenc}
\usepackage{amsmath}
\usepackage{amsfonts}
\usepackage{amssymb}
\usepackage{fancyhdr}
\usepackage{lastpage}
\usepackage{graphicx}
\usepackage{hyperref}
\usepackage{adjustbox}
\usepackage{verbatim}

\newcommand{\abs}[1]{\lvert#1\rvert}

\date{February 10, 2014}
\title{A Networks and Machine Learning Approach to Determine the Best College Coaches of the 20$^{\text{th}}$-21$^{\text{st}}$ Centuries}

\author{Tian-Shun Allan Jiang, Zachary T Polizzi, Christopher Qian Yuan \vspace{1mm}
		\and Mentor: Dr. Dan Teague\\
       The North Carolina School of Science and Mathematics
       \thanks{Paper submitted to the 2014 Mathematical Contest in Modeling (\url{http://www.comap.com/undergraduate/contests/mcm/})}
       }

\begin{document}

\maketitle
\newpage

\tableofcontents
\newpage

\section{Problem Statement}

College sport coaches often achieve widespread recognition. Coaches like Nick Saban in football and Mike Krzyzewski in basketball repeatedly lead their schools to national championships. Because coaches influence both the performance and reputation of the teams they lead, a question of great concern to universities, players, and fans alike is: Who is the best coach in a given sport? \textit{Sports Illustrated}, a magazine for sports enthusiasts, has asked us to find the best all-time college coaches for the previous century. We are tasked with creating a model that can be applied in general across both genders and all possible sports at the college-level. The solution proposed within this paper will offer an insight to these problems and will objectively determine the top five coaches of all time in the sports of baseball, men's basketball, and football.

\section{Planned Approach}

Our objective is to rank the top 5 coaches in each of 3 different college-level sports. We need to determine which metrics reflect most accurately the ranking of coaches within the last 100 years. To determine the most effective ranking system, we will proceed as follows:

\begin{enumerate}
\item Create a network-based model to visualize all college sports teams, the teams won/lost against, and the margin of win/loss. Each network describes the games of one sport over a single year. \label{itemnetwork}
\item Analyze various properties of the network in order to calculate the skill of each team.
\item Develop a means by which to decouple the effect of the coach from the team performance.
\item Create a model that, given the player and coach skills for every team, can predict the probability of the occurrence of a specific network of a) wins and losses and b) the point margin with which a win or loss occurred.
\item Utilize an optimization algorithm to maximize the probability that the coach skill matrix, once plugged into our model, generates the network of wins/losses and margins described in (\ref{itemnetwork}).
\item Analyze the results of the optimization algorithm for each year to determine an overall ranking for all coaches across history.
\end{enumerate}

\section{Assumptions}
Due to limited data about the coaching habits of all coaches at all teams over the last century in various collegiate sports, we use the following assumptions to complete our model. These simplifying assumptions will be used in our report and can be replaced with more reliable data when it becomes available.

\begin{itemize}

\item The skill level of a coach is ultimately expressed through his/her team's wins over another and the margin by which they win. This assumes that a team must win to a certain degree for their coach to be good. Even if the coach significantly amplifies the skills of his/her players, he/she still cannot be considered ``good" if the team wins no games.

\item The skills of teams are constant throughout any given year (ex: No players are injured in the middle of a season). This assumption will allow us to compare a team's games from any point in the season to any other point in the season. In reality, changing player skills throughout the season make it more difficult to determine the effect of the coach on a game.

\item Winning $k$ games against a good team improves team skill more than winning $k$ games against an average team. This assumption is intuitive and allows us to use the eigenvector centrality metric as a measure of total team skill.

\item The skill of a team is a function of the skill of the players and the skill of the coach. We assume that the skill of a coach is multiplicative over the skill of the players. That is: $T_s=C_s \cdot P_s$ where $T_s$ is the skill of the team, $C_s$ is the skill of the coach, and $P_s$ is a measure of the skill of the players. Making coach skill multiplicative over player skill assumes that the coach has the same effect on each player. This assumption is important because it simplifies the relationship between player and coach skill to a point where we can easily optimize coach skill vectors.

\item The effect of coach skill is only large when the difference between player skill is small. For example, if team $A$ has the best players in the conference and team $B$ has the worst, it is likely that even the best coach would not be able to, in the short run, bring about wins over team $A$. However, if two teams are similarly matched in players, a more-skilled coach will make advantageous plays that lead to his/her team winning more often than not.

\item When player skills between two teams are similarly matched, coach skill is the \textit{only} factor that determines the team that wins and the margin by which they win by. By making this assumption, we do not have to account for any other factors.

\end{itemize}

\section{Data Sources and Collection}

Since our model requires as an input the results of all the games played in a season of a particular sport, we first set out to collect this data. Since we were unable to identify a single resource that had all of the data that we required, we found a number of different websites, each with a portion of the requisite data. For each of these websites, we created a customized program to scrape the data from the relevant webpages. Once we gathered all the data from our sources, we processed it to standardize the formatting. We then aimed to merge the data gathered from each source into a useable format. For example, we gathered basketball game results from one source, and data identifying team coaches from another. To merge them and show the game data for a specific coach, we attempted to match on common fields (ex. ``Team Name"). Often, however, the data from each source did not match exactly (ex. ``Florida State" vs ``Florida St."). In these situations, we had to manually create a matching table that would allow our program to merge the data sources. 

Although we are seeking to identify the best college coach for each sport of interest for the last century, it should be noted that many current college sports did not exist a century ago. The National Collegiate Athletic Association (NCAA), the current managing body for nearly all college athletics, was only officially established in 1906 and the first NCAA national championship took place in 1921, 7 years short of a century ago. Although some college sports were independently managed before being brought into the NCAA, it is often difficult to gather accurate data for this time. 

\subsection{College Football}

One of the earliest college sports, College Football has been popular since its inception in the 1800's. The data that we collected ranges from 1869 to the present, and includes the results and final scores of every game played between Division 1 men's college football teams (or the equivalent before the inception of NCAA) \cite{shrpsports}. Additionally, we have gathered data listing the coach of each team for every year we have collected game data \cite{sportsref}, and combined the data in order to match the coach with his/her complete game record for every year that data was available.

\subsection{Men's College Basketball}

The data that we gathered for Men's College Basketball ranges from the season of the first NCAA Men's Basketball championship in 1939 to the present. Similarly to College Football, we gathered data on the result and final scores of each game in the season and in finals \cite{shrpsports}. Combining this with another source of coach names for each team and year generated the game record for each coach for each season \cite{sportsref}.

\subsection{College Baseball}

Although College Baseball has historically had limited popularity, interest in the sport has grown greatly in the past decades with improved media coverage and collegiate spending on the sport. The game result data that we collected ranges from 1949 to the present, and was merged with coach data for the same time period.


\section{Network-based Model for Team Ranking}


Through examination of all games played for a specific year we can accurately rank teams for that year. By creating a network of teams and games played, we can not only analyze the number of wins and losses each team had, but can also break down each win/loss with regard to the opponent's skill. 

\subsection{Building the Network}
We made use of a weighted digraph to represent all games played in a single year. Each node in the graph represents a single college sports team. If team $A$ wins over team $B$, a directed edge with a weight of 1 will be drawn from $A$ pointing towards $B$. Each additional time $A$ wins over $B$, the weight of the edge will be increased by 1. If $B$ beats $A$, an edge with the same information is drawn in the opposing direction. Additionally, a list containing the margin of win/loss for each game is associated with the edge. For example, if $A$ beat $B$ twice with $score: 64-60, 55-40$, an edge with weight two is constructed and the winning margin list {${4, 15}$} is associated with the edge. Since each graph represents a single season of a specific sport, and we are interested in analyzing a century of data about three different sports, we have created a program to automate the creation of the nearly 300 graphs used to model this system. The program \href{https://gephi.org/}{\textit{Gephi}} was used to visualize and manipulate the generated graphs. 

\begin{figure}[h]
    \begin{center}
       \includegraphics[angle=90, height=9cm]{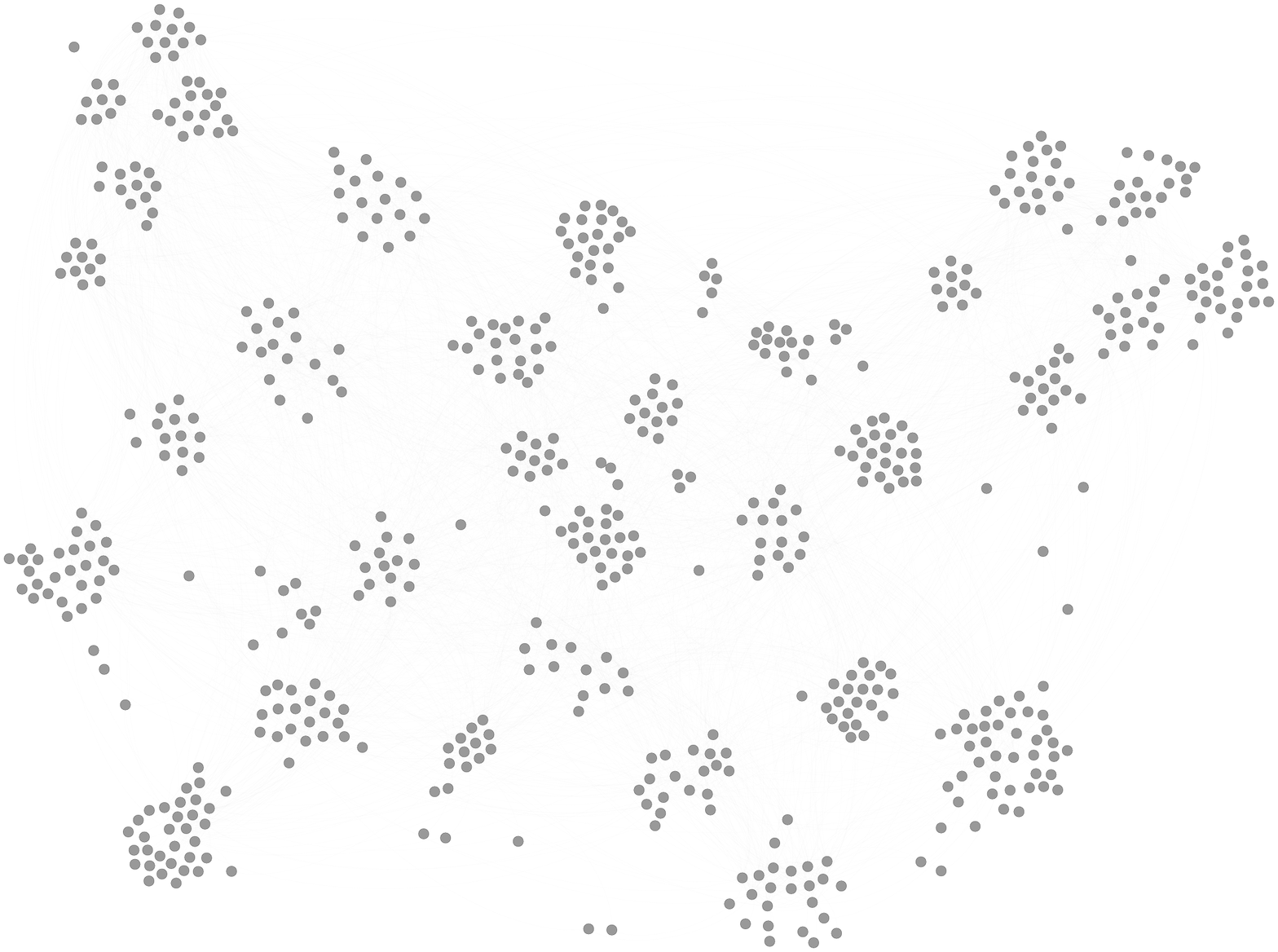}
        \caption{A complete network for the 2009-2010 NCAA Div. I basketball season. Each node represents a team, and each edge represents a game between the two teams. Note that, since teams play other teams in their conference most often, many teams have clustered into one of the 32 NCAA Div.1 Conferences.}
    \end{center}
\end{figure}
\subsection{Analyzing the Network}

We are next interested in calculating the skill of each team based on the graphs generated in the previous section. To do this, we will use the concept of centrality to investigate the properties of the nodes and their connections. Centrality is a measure of the relative importance of a specific node on a graph based on the connections to and from that node. There are a number of ways to calculate centrality, but the four main measures of centrality are degree, betweenness, closeness, and eigenvector centrality. 

\subsubsection{Degree Centrality}

Degree centrality is the simplest centrality measure, and is simply the total number of edges connecting to a specific node. For a directional graph, indegree is the number of edges directed into the node, while outdegree is the number of edges directed away from the node. Since in our network, edges directed inward are losses and edges directed outwards are wins, indegree represents the total number of losses and outdegree measures the total number of wins. Logically, therefore, $\frac{outdegreee}{indegreee}$ represents the $\frac{win}{loss}$ ratio of the team. This ratio is often used as a metric of the skill of a team; however, there are several weaknesses to this metric. The most prominent of these weaknesses arises from the fact that, since not every team plays every other team over the course of the season, some teams will naturally play more difficult teams while others will play less difficult teams. This is exaggerated by the fact that many college sports are arranged into conferences, with some conferences containing mostly highly-ranked teams and others containing mostly low-ranked teams. Therefore, win/loss percentage often exaggerates the skill of teams in weaker conferences while failing to highlight teams in more difficult conferences. 

\subsubsection{Betweenness and Closeness Centrality}

Betweenness centrality is defined as a measure of how often a specific node acts as a bridge along the shortest path between two other nodes in the graph. Although a very useful metric in, for example, social networks, betweenness centrality is less relevant in our graphs as the distance between nodes is based on the game schedule and conference layout, and not on team skill. Similarly, closeness centrality is a measure of the average distance of a specific node to another node in the graph - also not particularly relevant in our graphs because distance between nodes is not related to team skills.

\subsubsection{Eigenvector Centrality}

Eigenvector centrality is a measure of the influence of a node in a network based on its connections to other nodes. However, instead of each connection to another node having a fixed contribution to the centrality rating (e.g. degree centrality), the contribution of each connection in eigenvector centrality is proportional to the eigenvector centrality of the node being connected to. Therefore, connections to high-ranked nodes will have a greater influence on the ranking of a node than connections to low-ranking nodes. When applied to our graph, the metric of eigenvector centrality will assign a higher ranking to teams that win over other high-ranking teams, while winning over lower-ranking nodes has a lesser contribution. This is important because it addresses the main limitation over degree centrality or win/loss percentage, where winning over many low-ranked teams can give a team a high rank.

If we let $G$ represent a graph with nodes $N$, and let $A = (a_{n,t})$ be an adjacency matrix where $a_{n,t} = 1$ if node $n$ is connected to node $t$ and $a_{n,t} = 0$ otherwise. If we define $x_a$ as the eigenvector centrality score of node $a$, then the eigenvector centrality score of node $n$ is given by:

\begin{equation}
x_n = \frac{1}{\lambda} \sum_{t \in M(n)}x_t = \frac{1}{\lambda} \sum_{t \in G} a_{n,t}x_t
\end{equation}

where ${\lambda}$ represents a constant and $M(n)$ represents the set of neighbors of node $n$.

If we convert this equation into vector notation, we find that this equation is identical to the eigenvector equation:

\begin{equation}
\mathbf{Ax} = {\lambda}\mathbf{x}
\end{equation}

If we place the restriction that the ranking of each node must be positive, we find that there is a unique solution for the eigenvector $x$, where the $n^{th}$ component of $x$ represents the ranking of node $n$. There are multiple different methods of calculating $x$; most of them are iterative methods that converge on a final value of $x$ after numerous iterations. One interesting and intuitive method of calculating the eigenvector $x$ is highlighted below. It has been shown that the eigenvector $x$ is proportional to the row sums of a matrix $S$ formed by the following equation \cite{borgatti2005, spizzirri2011}:

\begin{equation}
S=A + \lambda^{-1}A^2 + \lambda^{-2}A^3 + ... + \lambda^{n-1}A^n + ...
\end{equation}

where $A$ is the adjacency matrix of the network and $\lambda$ is a constant (the principle eigenvalue). We know that the powers of an adjacency matrix describe the number of walks of a certain length from node to node. The power of the eigenvalue ($x$) describes some function of length. Therefore, $S$ and the eigenvector centrality matrix both describe the number of walks of all lengths weighted inversely by the length of the walk. This explanation is an intuitive way to describe the eigenvector centrality metric. We utilized NetworkX (a Python library) to calculate the eigenvector centrality measure for our sports game networks.

We can apply eigenvector centrality in the context of this problem because it takes into account both the number of wins and losses and whether those wins and losses were against ``good" or ``bad" teams. If we have the following graph: $A \rightarrow B \rightarrow C$ and know that $C$ is a good team, it follows that $A$ is also a good team because they beat a team who then went on to beat $C$. This is an example of the kind of interaction that the metric of eigenvector centrality takes into account. Calculating this metric over the entire yearly graph, we can create a list of teams ranked by eigenvector centrality that is quite accurate. Below is a table of top ranks from eigenvector centrality compared to the AP and USA Today polls for a random sample of our data, the 2009-2010 NCAA Division I Mens Basketball season. It shows that eigenvector centrality creates an accurate ranking of college basketball teams. The italicized entries are ones that appear in the top ten of both eigenvector centrality ranking and one of the AP and USA Today polls. 

\begin{center}
    \begin{tabular} { | c | c | c | c | }
    \hline
    \textbf{Rank} & \textbf{Eigenvector Centrality} & \textbf{AP Poll} & \textbf{USA Today Poll} \\ \hline
    1 & \textit{Duke} & \textit{Kansas} & \textit{Kansas} \\ \hline
    2 & \textit{West Virginia} & Michigan St. & Michigan St. \\ \hline
    3 & \textit{Kansas} & Texas & Texas \\ \hline
    4 & Syracuse & \textit{Kentucky} & North Carolina \\ \hline
    5 & \textit{Purdue} & Villanova & \textit{Kentucky} \\ \hline
    6 & Georgetown & North Carolina & Villanova \\ \hline
    7 & Ohio St. & \textit{Purdue} & \textit{Purdue} \\ \hline
    8 & Washington & \textit{West Virginia} & \textit{Duke} \\ \hline
    9 & \textit{Kentucky} & \textit{Duke} & \textit{West Virginia} \\ \hline
    10 & Kansas St. & Tennessee & Butler \\
    \hline
    
    \end{tabular}
    
\end{center}

As seen in the table above, six out of the top ten teams as determined by eigenvector centrality are also found on the top ten rankings list of popular polls such as AP and USA Today. We can see that the metric we have created using a networks-based model creates results that affirms the results of commonly-accepted rankings. Our team-ranking model has a clear, easy-to-understand basis in networks-based centrality measures and gives reasonably accurate results. It should be noted that we chose this approach to ranking teams over a much simpler approach such as simply gathering the AP rankings for various reasons, one of which is that there are not reliable sources of college sport ranking data that cover the entire history of the sports we are interested in. Therefore, by calculating the rankings ourselves, we can analyze a wider range of historical data.

Below is a graph that visualizes the eigenvector centrality values for all games played in the 2010-2011 NCAA Division I Mens Football tournament. Larger and darker nodes represent teams that have high eigenvector centrality values, while smaller and lighter nodes represent teams that have low eigenvector centrality values. The large nodes therefore represent the best teams in the 2010-2011 season.

\begin{figure}[h]
    \begin{center}
       \includegraphics[angle=0, height=11cm]{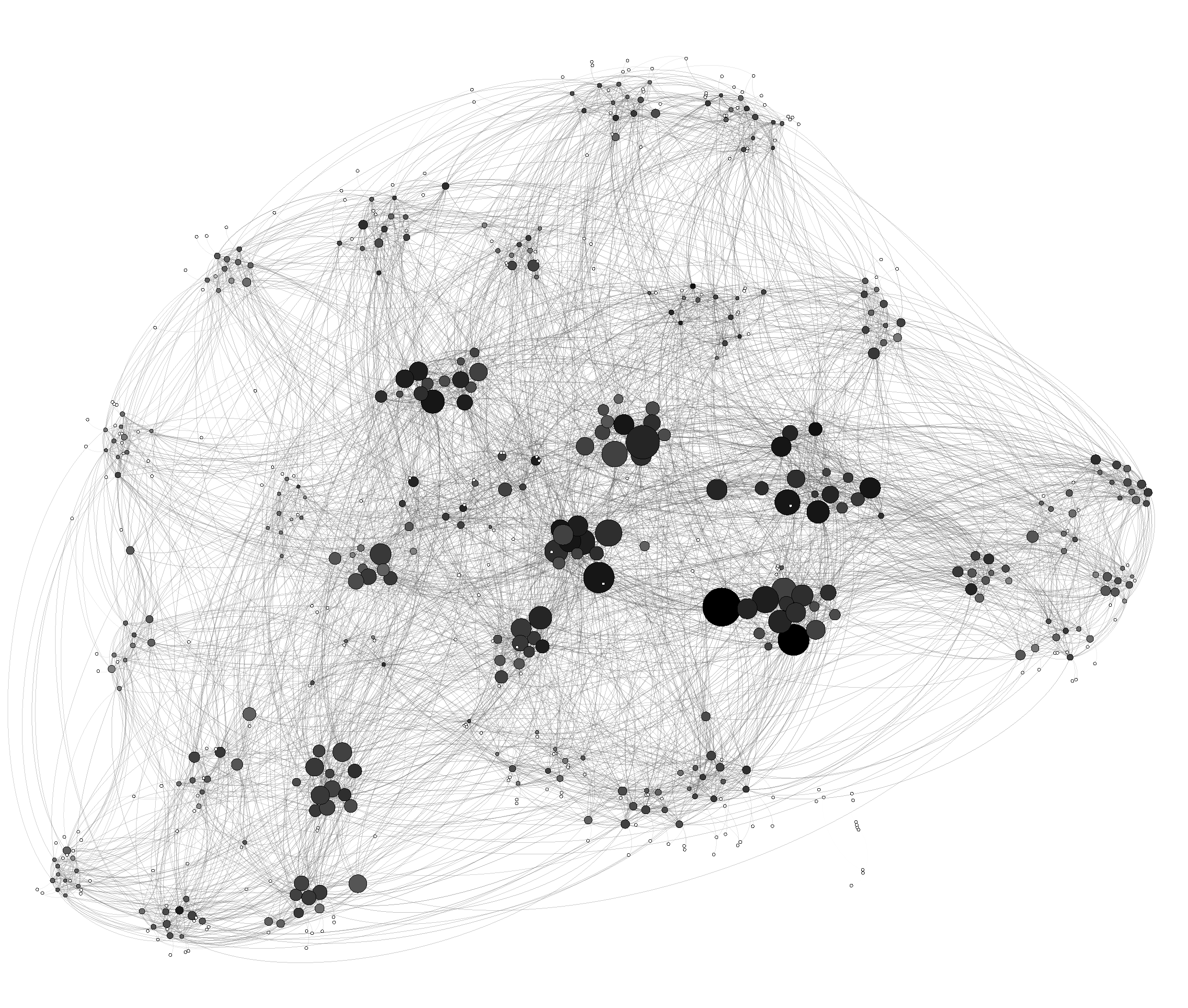}
        \caption{A complete network for the 2012-2013 NCAA Div. I Men's Basketball season. The size and darkness of each nodes represents its relative eigenvector centrality value. Again, note the clustering of teams into NCAA conferences.}
    \end{center}
\end{figure}

\section{Separating the Coach Effect}

The model we created in the previous section works well for finding the relative skills of teams for any given year. However, in order to rank the coaches, it is necessary to decouple the \textit{coach skill} from the overall team skill. Let us assume that the overall team skill is a function of two main factors, coach skill and player skill. Specifically, if $C_s$ is the coach skill, $P_s$ is the player skill, and $T_s$ is the team skill, we hypothesize that

\begin{equation}
T_s = C_s \cdot P_s, \label{2}
\end{equation}
as $C_s$ of any particular team could be thought of as a multiplier on the player skill $P_s$, which results in team skill $T_s$.

Although the relationship between these factors may be more complex in real life, this relationship gives us reasonable results and works well with our model.

\subsection{When is Coach Skill Important?}

We will now make a key assumption regarding player skill and coach skill. In order to separate the effects of these two factors on the overall team skill, we must define some difference in effect between the two. That is, the player skill will influence the team skill in some fundamentally different way from the coach skill.

\begin{figure}[h]
    \begin{center}
       \includegraphics[angle=0, height = 5cm]{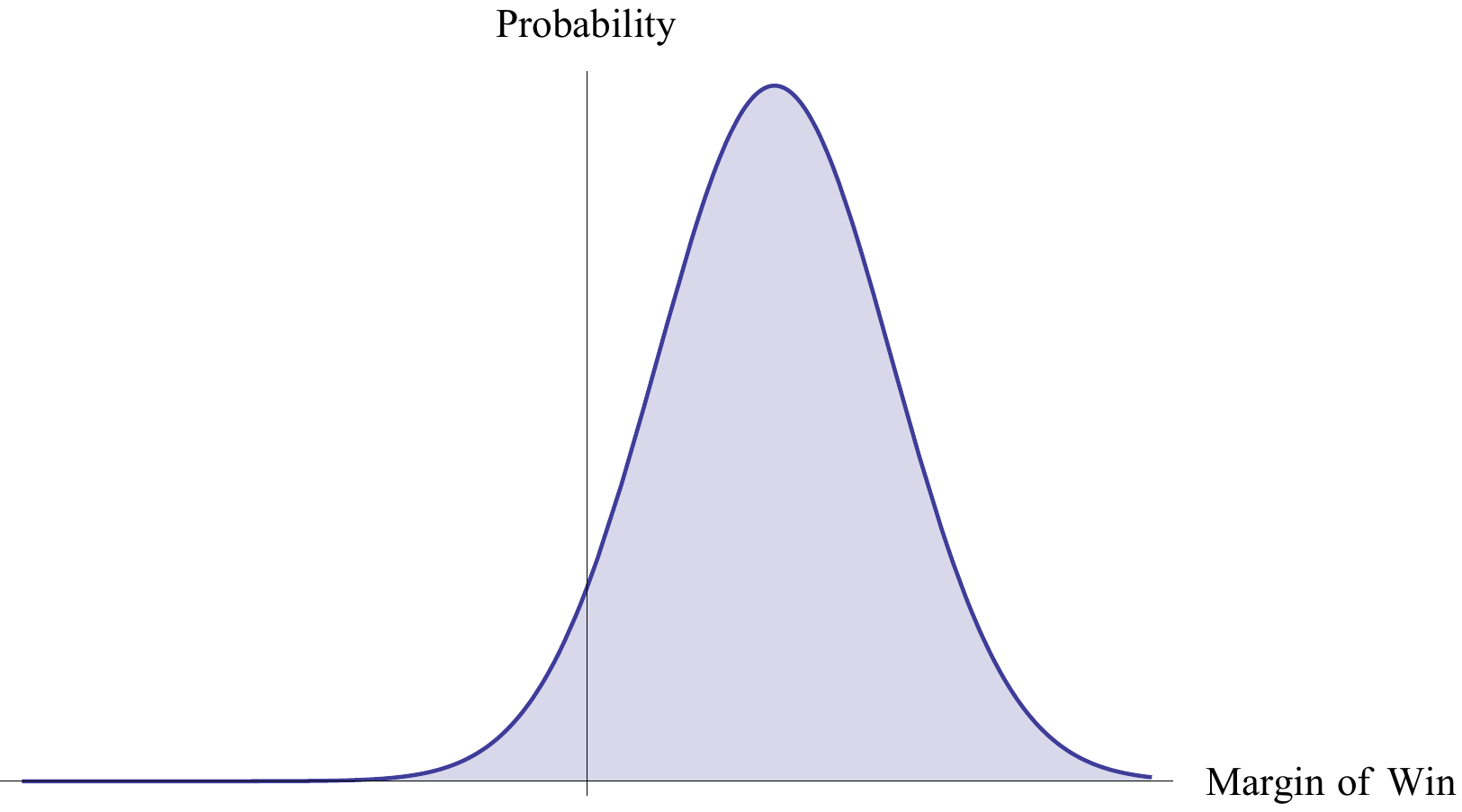}
        \caption{$A$ has a high chance of winning when its players are more skilled.} \label{mean2}
    \end{center}
\end{figure}

Think again to a game played between two arbitrary teams $A$ and $B$. There are two main cases to be considered:

\textbf{Case one: Player skills differ significantly:} Without loss of generality, assume that $P(A) >> P(B)$, where $P(x)$ is a function returning the player skills of any given team $x$. It is clear that $A$ winning the game is a likely outcome. We can draw a plot approximating the probability of winning by a certain margin, which is shown in Figure \ref{mean2}.

Because the player skills are very imbalanced, the coach skill will likely not change the outcome of the game. Even if $B$ has an excellent coach, the effect of the coach's skill will not be enough to make $B$'s win likely.

\begin{figure}[h]
    \begin{center}
       \includegraphics[angle=0, height=5cm]{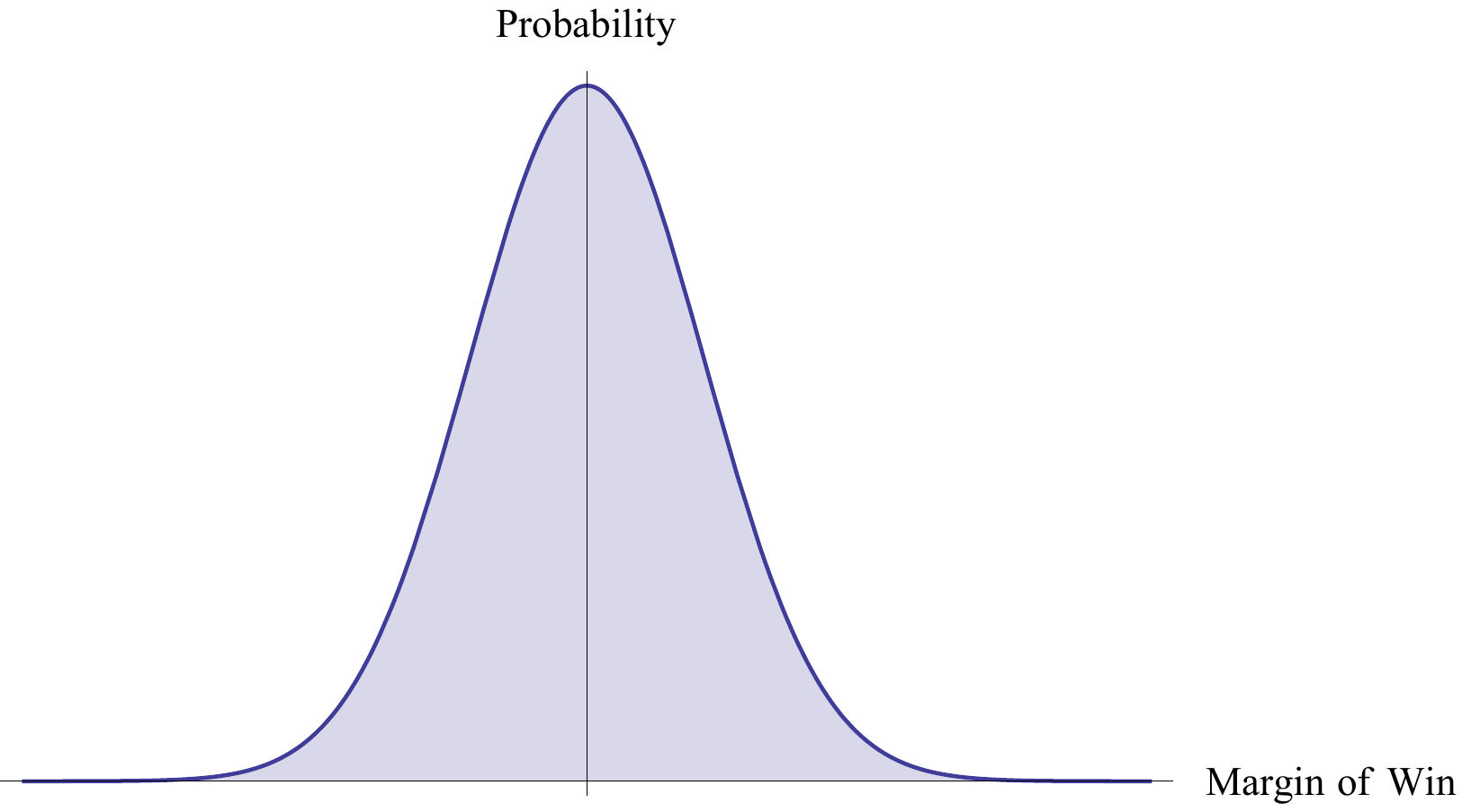}
        \caption{Neither $A$ nor $B$ are more likely to win when player skills are the same (if player skill is the only factor considered).} \label{mean0}
    \end{center}
\end{figure}

\textbf{Case two: Player skills approximately equal:} If the player skills of the two teams are approximately evenly matched, the coach skill has a much higher likelihood of impacting the outcome of the game. When the player skills are similar for both teams, the Gaussian curve looks like the one shown in Figure \ref{mean0}. In this situation, the coach has a much greater influene on the outcome of the game - crucial calls of time-outs, player substitutions, and strategies can make or break an otherwise evenly matched game. Therefore, if the coach skills are unequal, causing the Gaussian curve is shifted even slightly, one team will have a higher chance of winning (even if the margin of win will likely be small).

With the assumptions regarding the effect of coach skill given a difference in player skills, we can say that the effect of a coach can be expressed as:

\begin{equation}
(C_A - C_B) \cdot \left(\frac{1}{1+\alpha \abs{P_A-P_B}} \right) \label{coacheffect}
\end{equation}

Where $C_A$ is the coach skill of team $A$, $C_B$ is the coach skill of team $B$, $P_A$ is the player skill of team $A$, $P_B$ is the player skill of team $B$, and $\alpha$ is some scalar constant. With this expression, the coach effect is diminished if the difference in player skills is large, and coach effect is fully present when players have equal skill.

\subsection{Margin of Win Probability}

Now we wish to use the coach effect expression to create a function giving the probability that team $A$ will beat team $B$ by a margin of $x$ points. A negative value of $x$ means that team $B$ beat team $A$. The probability that $A$ beats $B$ by $x$ points is:

\begin{equation}
\displaystyle  K \cdot e^{\displaystyle -\left(\frac{1}{E}  \left(C \cdot \textit{player effect} + D \cdot \textit{coach effect} - \textit{margin} \right)
\right)^2} \label{marginprob}
\end{equation}
where $C, D, E$ are constant weights, \textit{player effect} is $P_A-P_B$, \textit{coach effect} is given by Equation \ref{coacheffect}, and \textit{margin} is $x$.

This probability is maximized when $$C \cdot \textit{player effect} + D \cdot \textit{coach effect} = \textit{margin}.$$ 
This accurately models our situation, as it is more likely that team $A$ wins by a margin equal to their combined coach and team effects over team $B$.

Since team skill is comprised of player skill and coach skill, we may calculate a given team's player skill using their team skill and coach skill.
Thus, the probability that team $A$ beats team $B$ by margin $x$ can be determined solely using the coach skills of the respective teams and their eigenvector centrality measures.

\subsection{Optimizing the Probability Function}

We want to assign all the coaches various skill levels to maximize the likelihood that the given historical game data occurred. To do this, we maximize the probability function described in Equation \ref{marginprob} over all games from historical data by finding an optimal value for the coach skill vectors $C_A$ and $C_B$. Formally, the probability that the historical data occurred in a given year is

\begin{equation}
\prod_{\text{all games}} K \cdot e^{\displaystyle -\left(\frac{1}{E}  \left(C \cdot \textit{player effect} + D \cdot \textit{coach effect} - \textit{margin} \right)\right)^2}.
\end{equation}

After some algebra, we notice that maximizing this value is equivalent to minimizing the value of the cost function $J$, where
\begin{equation}
J(C_s)
= \sum_{\text{all games}} (C \cdot \textit{player effect} + D \cdot \textit{coach effect} - \textit{margin})^2
\end{equation}

Because $P(A \text{ beats } B \text{ by } x )$ is a nonlinear function of four variables for each edge in our network, and because we must iterate over all edges, calculus and linear algebra techniques are not applicable. We will investigate three techniques (Genetic Algorithm, Nelder-Mead Search, and Powell Search) to find the global maximum of our probability function.

\subsubsection{Genetic Algorithm}
At first, our team set out to implement a Genetic Algorithm to create the coach skill and player skill vectors that would maximize the probability of the win/loss margins occurring. We created a program that would initialize 1000 random coach skill and player skill vectors. The probability function was calculated for each pair of vectors, and then the steps of the Genetic Algorithm were ran (carry over the ``most fit" solution to the next generation, cross random elements of the coach skill vectors with each other, and mutate a certain percentage of the data randomly). However, our genetic algorithm took a very long time to converge and did not produce the optimal values. Therefore, we decided to forgo optimization with genetic algorithm methods.

\subsubsection{Nelder-Mead Method}
We wanted to attempt optimization with a technique that would iterate over the function instead of mutating and crossing over. The Nelder-Mead method starts with a randomly initialized coach skills vector $C_s$ and uses a simplex to tweak the values of $C_s$ to improve the value of a function for the next iteration \cite{nelder1965}. However, running Nelder-Mead found local extrema which barely increased the probability of the historical data occurring, so we excluded it from this report. 

\subsubsection{Powell's Method}
A more efficient method of finding minima is Powell's Method. This algorithm works by initializing a random coach skills vector $C_s$, and uses bi-directional search methods along several search vectors to find the optimal coach skills. A detailed explanation of the mathematical basis for Powell's method can be found in Powell's paper on the algorithm \cite{powell1964}. We found that Powell's method was several times faster than the Nelder-Mead Method and produced reasonable results for the minimization of our probability function. Therefore, our team decided to use Powell's method as the main algorithm to determine the coach skills vector. We implemented this algorithm in Python and ran it across every edge in our network for each year that we had data. It significantly lowered our cost function $J$ over several thousand iterations.

\begin{center}
    \begin{tabular} { |c |  c| c | c|  }
    \hline
    Rank &  1962 & 2000 &2005 \\ \hline
    1&   John Wooden  & Lute Olson&Jim Boeheim\\ \hline
    2&   Forrest Twogood  &John Wooden &Roy Williams\\ \hline
    3&   LaDell Anderson  & Jerry Dunn &Thad Matta\\ \hline
 
    \hline
    
    \end{tabular}
    
\end{center}

The table above shows the results of running Powell's method until the probability function shown in Equation \ref{marginprob} is optimized, for three widely separated arbitrary years. We have chosen to show the top three coaches per year for the purposes of conciseness. We will additionally highlight the performance of our top three three outstanding coaches.

\textbf{John Wooden - UCLA:} John Wooden built one of the 'greatest dynasties in all of sports at UCLA', winning 10 NCAA Division I Basketball tournaments and leading an unmatched streak of seven tournaments in a row from 1967 to 1973 \cite{wooden}. He won 88 straight games during one stretch

\textbf{Jim Boeheim - Syracuse:} Boeheim has led Syracuse to the NCAA Tournament 28 of the 37 years that he has been coaching the team \cite{boeheim}. He is second only to Mike Krzyzewsky of Duke in total wins. He consistently performs even when his players vary - he is the only head coach in NCAA history to lead a school to four final four appearances in four separate decades.

\textbf{Roy Williams - North Carolina:} Williams is currently the head of the basketball program at North Carolina where he is sixth all-time in the NCAA for winning percentage \cite{williams}. He performs impressively no matter who his players are - he is one of two coaches in history to have led two different teams to the Final Four at least three times each.

\section{Ranking Coaches}

Knowing that we are only concerned with finding the top five coaches per sport, we decided to only consider the five highest-ranked coaches for each year. To calculate the overall ranking of a coach over all possible years, we considered the number of years coached and the frequency which the coach appeared in the yearly top five list. That is:

\begin{equation}
C_v=\frac{N_a}{N_c}
\end{equation}

Where $C_v$ is the overall value assigned to a certain coach, $N_a$ is the number of times a coach appears in yearly top five coach lists, and $N_c$ is the number of years that the coach has been active. This method of measuring overall coach skill is especially strong because we can account for instances where coaches change teams. 

\subsection{Top Coaches of the Last 100 Years}

After optimizing the coach skill vectors for each year, taking the top five, and ranking the coaches based on the number of times they appeared in the top five list, we arrived at the following table. This is our definitive ranking of the top five coaches for the last 100 years, and their associated career-history ranking:

\begin{center}
    \begin{tabular} { | c | c | c | c |  }
    \hline
    Rank & Mens Basketball & Mens Football & Mens Baseball \\ \hline
    1 & John Wooden - 0.28&Glenn Warner - 0.24& Mark Marquess - 0.27  \\ \hline
    2 & Lute Olson - 0.26& Bobby Bowden - 0.23& Augie Garrido - 0.24 \\ \hline
    3 & Jim Boeheim - 0.24& Jim Grobe - 0.18& Tom Chandler - 0.22 \\ \hline
    4 & Gregg Marshall - .23& Bob Stoops - 0.17& Richard Jones - 0.19 \\ \hline
    5 & Jamie Dixon - .21& Bill Peterson - 0.16& Bill Walkenbach - 0.16 \\ \hline

    \end{tabular}
    
\end{center}

\section{Testing our Model}
\subsection{Sensitivity Analysis}

A requirement of any good model is that it must be tolerant to a small amount of error in its inputs. In our model, possible sources of error could include improperly recorded game results, incorrect final scores, or entirely missing games. These sources of error could cause a badly written algorithm to return incorrect results. To test the sensitivity of our model to these sources of error, we decided to create intentional small sources of error in the data and compare the results to the original, unmodified results. 

The first intentional source of error that we incorporated into our model was the deletion of a game, specifically a regular-season win for Alabama (the team with the top-ranked coach in 1975) over Providence with a score of 67 to 60. We expected that the skill value of the coach of the Alabama team would decrease slightly with this modification. When we ran and analyzed the results, we found that the coach skill value did in fact decrease by approximately 1\%, as we expected. However, the Alabama coach maintained his ranking of top coach for the season. 

The second change that we incorporated was to switch the results of the same game (Alabama 67, Providence 60) to a win for Providence (Providence 67, Alabama 60). We expect this will have a greater negative influence on the skill value of the Alabama coach, and when we ran the analysis we found that, indeed, the Alabama coach skill value decreased by approximately 4\%. Although a relatively minor difference, the second-ranked coach originally had a skill value only very slightly behind the Alabama coach, and the 4\% loss in fact placed the second-ranked coach in the first ranking position.

From this analysis we can see that our model follows our predictions accurately, and that removing factors that add positively to the skill ranking of a coach is detrimental to their skill value. Although the changes we made were minor, there is often a lot of competition for the first-place ranking of a coach, and due to the limited number of games played per season, a change to this data can have an influence on the final ranking. Although these results indicate that error in our data can effect the final ranking, the analysis also shows that our model responds predictably to a variation in the input.

\subsection{Strengths}
The main strength of our approach is that it is able to separate coach proficiency from team proficiency by calculating probabilities that the historical game data occur given coach skills. This allows us to more accurately gauge the skills of a coach without factoring in the skills of his/her players. Furthermore, our approach is flexible as many relationships can be modified. For example, if a study shows that there is a better function to describe the relationship between coach skill, player skill, and team skill, it can easily be used in our model. Our model is also able to compare the relative effectiveness of coaches from all time periods, as long as the average margin of victory is similar across time periods.

\subsection{Weaknesses}
The main weakness of our model pertains to computational efficiency, as our computers could not always adequately calculate all necessary values in our model. For example, the computer could not find the eigenvector centrality values on a small percentage of the graphs, as the Von Mises iteration failed to converge. Furthermore, sometimes Powell's method of minimizing our cost function yielded high costs relative to other years because the initialized array of coach skills was close to a local minima. This could be solved by running Powell's method from several randomly initialized coach skills arrays, but this increases computational time. In fact, the overall results of our model could likely be improved significantly given more time to run the iterative optimization algorithms with a higher accuracy, resulting in better approximations for the ideal matrix of coach skills.

\section{Conclusions}

In this report, we have analyzed nearly 300 years of data in order to determine the most accurate and unbiased ranking of all-time best college sports coaches. By constructing comprehensive networks with with edges representing each and every game played in the last century of the college sports that we analyzed, we were able to create a comprehensive metric of team skill using the concept of eigenvector centrality. By considering win/loss margins, we were able to identify patterns that enabled us to separate the team skill measure into its two components - player skill and coach skill. We then created a probability function based on player skill and coach skill to determine the likelihood of an edge in our network occurring. By multiplying this probability across all edges, we were able to determine the probability of the entire graph occurring given team skill and coach skill vectors. Using an iterative, multivariable, machine learning algorithm, we maximized this probability function for coach skill for each season and for each sport. Using data that mapped the name of a coach to his/her team for each season, we were able to combine the results of each individual season and analyze the skill of each individual coach over their entire coaching history. From this data, we selected the top 5 coaches from every sport to feature as our all-time best coaches of the century.

\section{Acknowledgments}
We would like to thank our mentor Dr. Dan Teague, who introduced us to the world of mathematical modeling, and Ms. Christine Belledin, who sponsored us for MCM. We would also like to thank the North Carolina School of Science and Mathematics for hosting us as we worked on the problem. Last but not least, we would like to thank our judges and Comap for putting together this wonderful competition.

\newpage

\begin{thebibliography}{1}

\bibitem{wooden}
{John Wooden}.
 Retrieved from:
  \url{http://msn.foxsports.com/collegebasketball/story/John-Wooden-dies-UCLA-coach-99-060410},
  2010.

\bibitem{shrpsports}
{ShrpSports}.
 Retrieved from: \url{http://www.shrpsports.com/}, 2011.

\bibitem{boeheim}
{Jim Boeheim}.
 Retrieved from:
  \url{http://cuse.com/coaches.aspx?rc=405&path=mbasket}, 2012.

\bibitem{sportsref}
{Sports-Reference}.
 Retrieved from: \url{http://www.sports-reference.com/}, 2013.

\bibitem{williams}
{Roy Williams}.
 Retrieved from:
  \url{http://www.goheels.com/ViewArticle.dbml?ATCLID=205497516}, 2014.

\bibitem{borgatti2005}
Stephen~P. Borgatti.
 {Centrality and network flow}.
 {\em Social Networks}, 27(1):55--71, January 2005.

\bibitem{nelder1965}
JA~Nelder and R~Mead.
 {A simplex method for function minimization}.
 {\em The Computer Journal}, 1965.

\bibitem{powell1964}
MJD Powell.
 {An efficient method for finding the minimum of a function of several
  variables without calculating derivatives}.
 {\em The Computer Journal}, 1964.

\bibitem{spizzirri2011}
Leo Spizzirri.
 {Justification and Application of Eigenvector Centrality}.
 2011.

\end{thebibliography}
\bibliographystyle{plain}

\end{document}